\documentclass[12pt]{amsart}
\DeclareUnicodeCharacter{2212}{-}
\usepackage{amsmath,amsthm,amsfonts,amscd,amssymb,mathrsfs}
\usepackage{stmaryrd}
\usepackage{mathtools}
\usepackage[T1]{fontenc}
\usepackage[utf8]{inputenc}
\usepackage{geometry} 
\usepackage{graphicx}
\usepackage{float}
\usepackage{physics}
\usepackage{nccmath}
\usepackage{url}
\usepackage{caption}
\usepackage{cite}
\usepackage[initials]{amsrefs}
\usepackage{placeins}
\usepackage{booktabs} 
\usepackage{epstopdf}
\usepackage{soul} 
\usepackage{nccmath} 
\usepackage{url}
\usepackage{hyperref}
  \hypersetup{colorlinks=true,citecolor=blue, urlcolor= cyan}
\usepackage{cleveref}
 \AtBeginDocument{%
    \def\MR#1{} 
 }
\usepackage{color}

\usepackage{epsfig}

\usepackage{array} 
\usepackage{verbatim}

\usepackage{longtable}
\usepackage{extarrows}
\usepackage{xcolor}
\usepackage{qcircuit}

\newcommand{\CC}{\mathbb{C}}

\title{Four-Qubit CHSH Games}

\author{Joaquim Jusseau}\email{joaquim.jusseau@utbm.fr}\address{University of Technology of Belfort-Montbéliard, 90010 Belfort Cedex, France}
\author{Hamza Jaffali}\email{hamza.jaffali@colibri.com}\address{ColibrITD, 91 Rue du Faubourg Saint Honor\'e, 75008 Paris, France}
\author{Fr\'ed\'eric Holweck}\email{frederic.holweck@utbm.fr}\address{Laboratoire Interdisciplinaire Carnot de Bourgogne,  UMR 6303 CNRS, University of Technology of Belfort-Montbéliard, 90010 Belfort Cedex, France}\address{Mathematics and Statistics Department, Auburn University, Auburn, AL, USA}

\date{}

\begin{document}
\begin{abstract}
In this paper, the CHSH quantum game is extended to four players. This is achieved by exploring all possible 4-variable Boolean functions to identify those that yield a game scenario with a quantum advantage using a specific entangled state. Notably, two new four-player quantum games are presented. In one game, the optimal quantum strategy is achieved when players share a $\ket{GHZ}=\dfrac{1}{\sqrt{2}}(\ket{0000}+\ket{1111})$ state, breaking the traditional 10\% gain observed in 2 and 3 qubit CHSH games and achieving a 22.5\% gap. In the other game, players gain a greater advantage using a $\ket{W}=\dfrac{1}{2}(\ket{0001}+\ket{0010}+\ket{0100}+\ket{1000})$ state as their quantum resource. Quantum games with other four-qubit entangled states are also explored. To demonstrate the results, these game scenarios are implemented on an online quantum computer, and the advantage of the respective quantum resource for each game is experimentally verified.
\end{abstract}

\maketitle

\section{Introduction}

Quantum games are games where players are allowed to use quantum resources like entanglement or quantum contextuality to establish winning strategies that outperform their classical counterparts. These games explore, within game scenarios, some paradoxical properties of quantum physics \cites{aravind2004quantum, cabello2001all, kunkri2007winning, renner2004quantum}. One of the most famous games in the field is the CHSH game, named after Clauser, Horne, Shimony, and Holt \cite{clauser1969proposed}, which paved the way for the realization of the famous experiments of Aspect \cite{aspect1982experimental} on the $EPR$ paradox and the non-locality of quantum physics. Its game-theoretic version is the focus of this paper.

Let's consider the principle of the two-qubit CHSH game: a referee Charlie ($C$) is playing with Alice ($A$), and Bob ($B$). During the game, Alice and Bob are prohibited from communicating. Charlie sends each player a binary question: Alice receives $x$ and Bob receives $y$, where \(x, y \in \mathbb{B} = \{0, 1\}\). Each player must respond with a binary answer that satisfies a specific equation. Alice responds with bit $a$, and Bob responds with bit $b$, without knowledge of the other's response. In the traditional CHSH game, the players win by providing answers $(a,b)$ to the questions $(x,y)$ that fulfill the equation:

\begin{equation}\label{eq:original}
    x \cdot y = a \oplus b ~.
\end{equation}

Alice and Bob can agree on a strategy beforehand, but with classical resources and no communication during the game\footnote{These conditions are referred to as Local Realism (LR). When using Quantum Mechanics (QM) rules instead, it is denoted by QM.}, their maximum winning probability is $p_{LR} \leq 0.75$. If they utilize quantum resources, they can prepare a so-called $EPR$ state, $\ket{EPR}=\dfrac{\ket{00}+\ket{11}}{\sqrt{2}}$, share the qubits (one for Alice, one for Bob), and then apply the following strategy:

\begin{itemize}
\item If $x=0$, Alice measures her qubit in the $Z$-basis (standard basis) and returns $0$ for a $+1$ measurement and $1$ otherwise.
\item If $x=1$, Alice measures her qubit in the $X$-basis (Hadamard basis) and returns $0$ for a $+1$ measurement and $1$ otherwise.
\item If $y=0$, Bob measures his qubit in the $\dfrac{Z+X}{\sqrt{2}}$ basis and returns $0$ for a $+1$ measurement and $1$ otherwise.
\item If $y=1$, Bob measures his qubit in the $\dfrac{Z-X}{\sqrt{2}}$ basis and returns $0$ for a $+1$ measurement and $1$ otherwise.
\end{itemize}

In \cite{Jaffali2024} the CHSH game was extended to three players and one referee by exploring the space of quantum strategies that allow the three players to win the game with a better rate of success than any classical strategy. In this generalization the winning condition defining  a game is defined by a boolean equation:

\begin{equation}\label{eq:2}
f(a,b,c)=g(x,y,z) \ \text{mod } 2 ~.
\end{equation}

All games given by an equation of the form of Eq. (\ref{eq:2})  were investigated in \cite{Jaffali2024} and one of the outcomes of this work was that different types of entangled states, known as the $GHZ$ and $W$ states, could be advantageously used for different games.

The complexity in generalizing this approach to the $n=4$ is of two kinds. First, the number of boolean equations to consider increases dramatically (see Section \ref{sec:generalization}). Then the number of quantum resources, i.e. type of entangled states, is infinite in the four-qubit case \cite{verstraete2002four}. 
Those difficulties led us to adapt the method of \cite{Jaffali2024} but also led to new interesting results. In particular one found a four-qubit game scenario where the quantum strategy increases by more than $22\%$ the chance of winning the game compared to any classical strategy. Compared to the $10\%$ improvement of the quantum strategy in the CHSH game or in all three-qubit games of \cite{Jaffali2024}, it is a notable gap that we confirmed by some experiments on an online quantum computer. 

The structure of this paper is as follows. Section \ref{sec:generalization} recalls the method to generalize the CHSH game. The approach involves considering all possible binary equations of type $f(x_1,\dots, x_n)=g(a_1,\dots, a_n)$ with $2n$ binary variables, where the first $n$ variables form the set of questions  $(x_1,\dots, x_n) \in \mathbb{B}^n$, and the $n$ variables $(a_1,\dots ,a_n)\in \mathbb{B}^n$ represent the set of possible answers provided by the $n$ players. Given an entangled state as a shared resource among the players, one can test whether a quantum advantage exists for a given equation by exploring all possible quantum strategies with that specific resource. For $n=4$, the methods of selection used to reduce the problem's size are explained.

In Section \ref{sec:4qgame}, the setup of the four-qubit game is presented, and a new type of equation (game) is identified where the four players have a probability of winning with a quantum strategy based on the $\ket{GHZ}$ state that largely outperforms the best classical one. The search is also conducted when the resource is a $\ket{W}$ state, and an example of games where the use of a strategy based on $\ket{W}$ beats the classical strategy and the quantum strategy obtained with a $\ket{GHZ}$ state is provided. Other types of states are tested and discussed including some maximally entangled states as well as families of four-qubit quantum states provided by the  classification in \cite{verstraete2002four}.


In Section \ref{sec:ibm}, implementation on the IBM Quantum Platform of two new games is described: one showing an advantage using a $\ket{GHZ}$ state as a resource for the four players, and one based on $\ket{W}$. For each game, it is shown that the experimental results produced by the online quantum computer beat the classical bound and allow distinguishing the two resources according to the outcomes of the measurement. 

Finally, Section \ref{sec:conclusion} is dedicated to concluding remarks and possible extensions of the work.

\section{Generalization of the CHSH game} \label{sec:generalization}
In this subsection, the original CHSH game is adapted to create a generalized version for $n$ players. The approach involves listing all possible binary equations for a given $n$ and comparing the probabilities of winning the game using classical strategies versus a quantum strategy based on shared $n$-qubit entangled states. The method for evaluating and determining both classical and quantum strategies for various setups of the generalized CHSH game is also detailed. 

\subsection{Definition of the game}

As in \cite{Jaffali2024}, the $n$ players are denoted by $A_1, A_2, \dots, A_n$. Each player $i \in \llbracket 1, n \rrbracket$ receives a question $x_i \in \mathbb{B}$ from the referee and must provide an answer $a_i \in \mathbb{B}$. To win the game, the players need to satisfy the following general equation:

\begin{equation}
f(x_1, x_2, \dots, x_n) = g(a_1, a_2, \dots, a_n) ~.
\end{equation}

Here, $f$ and $g$ are boolean functions from $\mathbb{B}^n$ to $\mathbb{B}$, ensuring that each term $x_i$ and $a_i$ appears at least once in the expressions for $f$ and $g$, respectively, to ensure every player is fully involved in the game.

The strategy that the players will use to win the game is denoted by $h$, defined as follows:

\begin{equation}
h \colon \biggl\{\begin{array}{@{}r@{\;}l@{}}
    \mathbb{B}^n  & \to \mathbb{B}^n \\
    (x_1, x_2, \dots, x_n) & \mapsto (a_1, a_2, \dots, a_n)
  \end{array} ~,
\end{equation}

which specifies the answers the players will give based on the questions they receive.

In other words, the players aim to find a strategy $h$ that maximizes the satisfaction of the equation:

\begin{equation}
f(x_1, x_2, \dots, x_n) = g(h(x_1, x_2, \dots, x_n)) ~.
\end{equation}

\subsection{Finding the best classical strategy}

For a classical deterministic strategy, each player will decide in advance the answer to give to the question they receive. Thus, the strategy $h$ can be decomposed as follows:

\begin{equation}
h(x_1, x_2, \dots, x_n) = \big(h_1(x_1), h_2(x_2), \dots, h_n(x_n) \big)
\end{equation}

where $h_i$ represents the strategy for the $i$-th player. For each player $i$, defining a deterministic strategy involves specifying $h_i(0)$ and $h_i(1)$. Consequently, $2n$ bits are required to encode any $n$-player strategy $h$, leading to a total of $2^{2n}$ possible classical strategies. To find the optimal strategy, all possible strategies are generated, each is evaluated for its probability of winning, and the strategies that yield the best performance are identified.

\subsection{Finding the best quantum strategy}

For a quantum strategy, the players share an $n$-qubit quantum state $\ket{\psi_n}$. To harness the advantages of quantum mechanics, $\ket{\psi_n}$ must be an entangled state; otherwise, it reduces to a stochastic classical strategy. Each player can manipulate their qubit by applying unitary operations. Depending on the question received, a player will apply one of two possible operations defined by their strategy. Let $U_{i,x_i}$ denote the unitary gate applied by player $i$ on their qubit when they receive question $x_i$. After all players have applied their strategies, they measure their qubits and provide the results of the measurements as their answers. This process is summarized in Figure \ref{fig}.

\begin{figure}[!h]
\begin{equation*}
(x_1, x_2, \dots, x_n) ~ \xlongrightarrow{\text{apply strategy}} ~ U_{1,x_1} \otimes \cdots \otimes U_{n,x_n} \ket{\psi_n} ~ \xlongrightarrow{\text{measure}} ~ (a_1, a_2, \dots, a_n)
\end{equation*}
\caption{Overview of the application of an $n$-player quantum strategy.}
\label{fig}
\end{figure}

As written in \cite{Jaffali2024}, any unitary operator $U_{i,x_i}$ acting on one qubit can be parameterized by three angles $\theta_{i,x_i}$, $\phi_{i,x_i}$ and $\lambda_{i,x_i}$, such that:

\begin{equation}
    U_{i,x_i}(\theta_{i,x_i}, \phi_{i,x_i},\lambda_{i,x_i}) = 
    \begin{pmatrix}
    \cos\left(\frac{\theta_{i,x_i}}{2}\right) &  -e^{j\lambda_{i,x_i}} \sin\left(\frac{\theta_{i,x_i}}{2}\right) \\
    e^{j\phi_{i,x_i}}  \sin\left(\frac{\theta_{i,x_i}}{2}\right) &        e^{j(\phi_{i,x_i} + \lambda_{i,x_i})} \cos\left(\frac{\theta_{i,x_i}}{2}\right)
    \end{pmatrix}
\end{equation}

with $j$ the complex imaginary unit.

\subsection{Problem's reduction}
Recall that there are \(2^{16} = 65,536\) possible functions from \(\mathbb{B}^4\) to \(\mathbb{B}\). Therefore, an exhaustive analysis would require testing \(65,536^2\) equations in order to analyze all games defined by a winning condition of type $f(w,x,y,z)=g(a,b,c,d)$.
Finding the optimal angles $\theta_{i,x_i}$, $\phi_{i,x_i}$, and $\lambda_{i,x_i}$ for a single equation takes a minimum of 2 seconds with Python on a regular Laptop. In order to enhance execution time and optimize functions, as well as to reduce the number of functions processed while covering the maximum possible cases, various techniques  are used. These techniques include optimizing algorithms, parallel processing, and implementing selection methods. But first, we simplify the setting by fixing a choice of a function $g$ and reducing the analysis to the $65, 536$ possibles $f$.

\subsubsection{Functions reduction} \label{sec:method}
The first step is to divide by 2 the number of functions to process. Indeed one can remark that the game defined by $\overline{f(w,x,y,z)}=g(a,b,c,d)$ and the game defined by $f(w,x,y,z)=g(a,b,c,d)$, have the same optimal classical and quantum gains if $g$ has symmetries that allows one player to flip the result of $g(a,b,c,d)$ by changing his/her answer. For instance if $g(a,b,c,d)=a\oplus b\oplus c\oplus d$ any wining strategy for $f(w,x,y,z)=g(a,b,c,d)$ is a winning strategy for $\overline{f(w,x,y,z)}=g(a,b,c,d)$ if a unique player flip his/her answer. Thus it reduces the number of functions to process  to $32,768$ if one restricts to this type of function $g$.
The second step involves eliminating variants of the same function. For a given function $f$ with $n$ variables, we can generate $2^n$ possible variants by applying NOT gates to the $n$ variables which correspond to equivalent games. After this step, the number of functions to process is reduced to $4,014$ for $n=4$. The final step is to select only the function letting all game variables $w,x,y,z$ and $a,b,c,d$ appear in the equation and after this final the number of function to test is $3907$. \\
After those reductions, the final execution time to test 3907 functions $f(w,x,y,z)$ for one given function $g(a,b,c,d)$ is around 40 minutes.

\subsubsection{Parallel processing}
To parallelize the optimization function and utilize all CPU cores effectively, the \texttt{joblib} library was utilized. This enabled simultaneous optimization processes across multiple CPU cores, resulting in a significant reduction in overall execution time.

\section{The four-qubit CHSH games}\label{sec:4qgame}
In this section,  new and non-equivalent game versions of the CHSH games for four players are investigated. We first consider the games based on quantum strategies using as four-qubit entangled states either a $\ket{GHZ}$ or a $\ket{W}$ state. Recall that:
\begin{itemize}
    \item $\ket{GHZ}=\dfrac{1}{\sqrt{2}}(\ket{0000}+\ket{1111})$,
    \item $\ket{W}=\dfrac{1}{2}(\ket{1000}+\ket{0100}+\ket{0010}+\ket{0001})$.
\end{itemize}
Then we also discuss other examples of four-qubit genuine entangled states known as critical states and finally one considers the full four-qubit classification as described in \cite{verstraete2002four}.

\subsection{Game's setup}
The four players will be called $J1, J2, J3$  and $J4$. Here, as explained in the general case in Section \ref{sec:generalization}, the four players play the same game, share a unique four-qubit state, and try to satisfy the same equation.

\begin{figure}[!h] 
    \centering
    \includegraphics[width=0.8\linewidth]{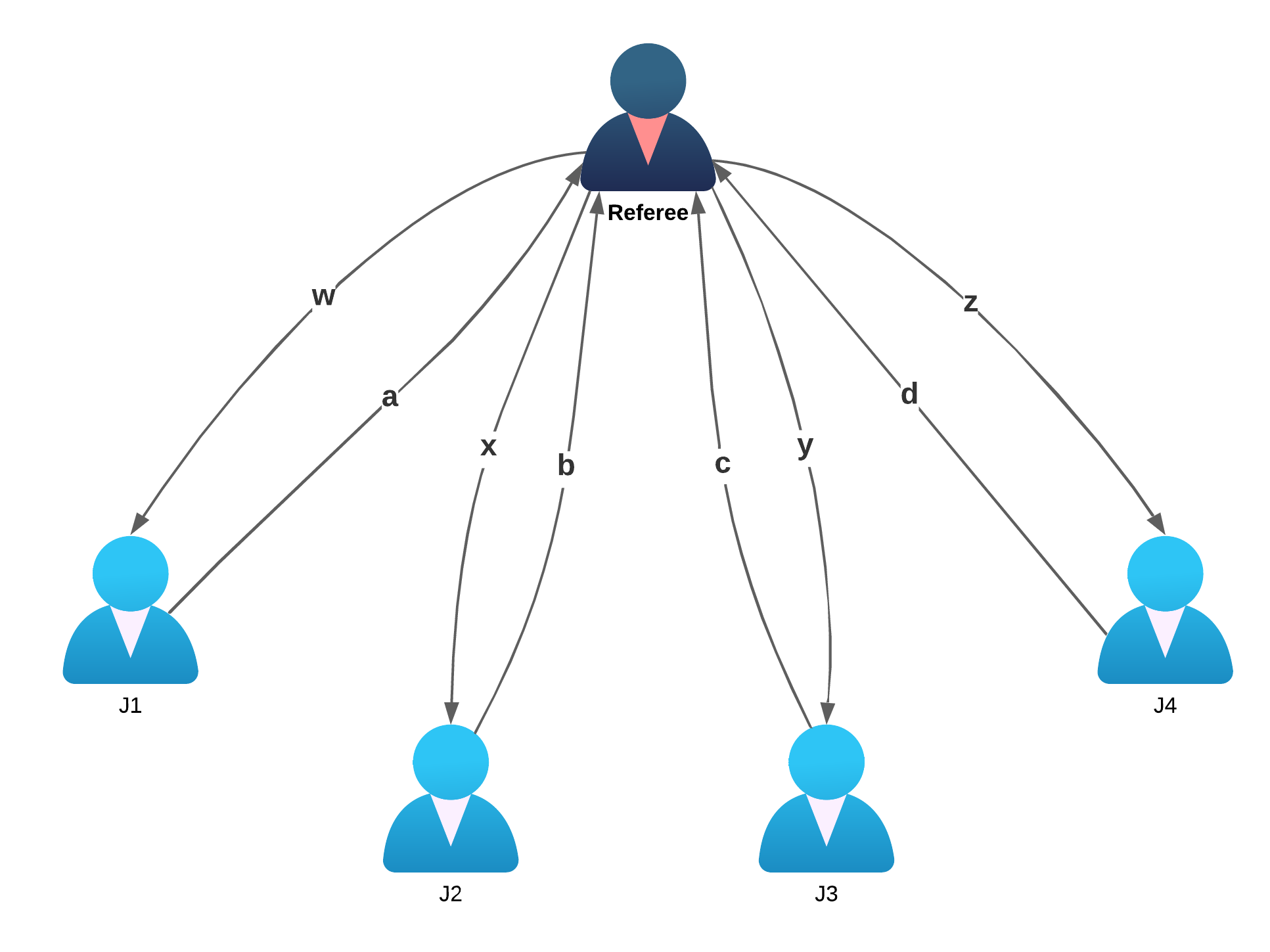}
    \caption{ CHSH game setup with 4 players: The referee sends a question $w$ to $J1$, $x$
to $J2$, $y$ to $J3$ and $z$ to $J4$. The players win the game if and only if their answers $(a, b, c,d)$
satisfy the binary equation $f(w, x, y,z) = g(a, b, c,d)$ that defines the game.}
    \label{fig:game_setup}
\end{figure}
The referee sends respectively the binary questions $w$, $x$, $y$, and $z$, to $J1, J2, J3$ and $J4$, they apply their strategy, and they answer respectively $a$, $b$, $c$ and $d$ (see Figure \ref{fig:game_setup}). The gain is then deduced from the satisfaction of the equation of the game, which should take the following form:

\begin{equation}\label{eq:3q-problem}
    f(w,x,y,z) = g(a,b,c,d)
\end{equation}

As recalled in Section \ref{sec:method}, it is necessary to filter the useful equations from all the possible ones. We listed the 3907 possible functions from $\mathbb{B}^4$ to $\mathbb{B}$, after removing the non-relevant ones.

\subsection{Gain with $\ket{GHZ}$}
Concerning the choice of the $g$ function, we first generalized the function $g(a,b) =  a \oplus b$ used in the 2-qubit CHSH game (see Eq. (\ref{eq:original})). For four players, it becomes: 
$$g(a,b,c,d) = a \oplus b \oplus c \oplus d$$

The first interesting fact is that for four-qubit CHSH games it is possible to have a gap between quantum and classical strategies higher than 10\% (the limit observed in the two and three-qubit CHSH game \cite{Jaffali2024}), but indeed a gap of 22.5\% is achieved for the game defined by:

\begin{equation}\label{eq:GHZ}
    (x y  z) + (x  y \overline{w}) + (x  z  \overline{w}) + (y  z  \overline{w} ) + (w  \overline{x} \overline{y} \overline{z}) = a \oplus b \oplus c \oplus d 
\end{equation}

This good performance can be attributed to the symmetric nature of the function $g(a,b,c,d) = a \oplus b \oplus c \oplus d $ and the strong correlations and entanglement properties inherent in the $\ket{GHZ}$ state. This symmetry aligns well with the requirements of the equation, ensuring that the outcomes of the quantum measurements match the desired function.
An explicit quantum strategy is given in Table \ref{tab:examples_GHZ_strategies}. Note that the quantum gain is of $85.35\%$ which is similar to the quantum gains obtained for the analogue two and three-qubit CHSH games with the best strategy. But the gap between quantum and classical is much larger due to the impossibility to win the game classically with a probability higher than $0.6225$ in the four players game. The larger number of variables (players) makes it more difficult to find a deterministic strategy.

\begin{table}[h!] 
    \centering
    \begin{tabular}{|c|c|c|}
        \hline 
        Equation  & $U_{1,0}$ & $U_{1,1}$ \\
        \cline{1-2} 
        \hline
         & $(\frac{3\pi}{2}, 2.7153, 4.4219)$ & $(\frac{\pi}{2}, 4.9531, 5.9927)$\\
        \cline{2-3}
        \cline{2-3}
        &  $U_{2,0}$ & $U_{2,1}$ \\
        \cline{2-3} 
        $(x y  z) + (x  y \overline{w}) + (x  z  \overline{w})+ (y  z  \overline{w} )$  &  $(\frac{3\pi}{2}, 3.7575,4.9831)$ & $(\frac{3\pi}{2}, 3.9628, 0.2707)$ \\
        \cline{2-3} 
      $ + (w  \overline{x} \overline{y} \overline{z})= a \oplus b \oplus c \oplus d$ & $U_{3,0}$ & $U_{3,1}$ \\
        \cline{2-3} 
           & $(\frac{\pi}{2}, 6.0502, 3.6010)$ & $(\frac{\pi}{2}, 6.0234, 5.1718)$ \\
        \cline{2-3}
        \cline{2-3} 
        \cline{2-3} 
        & $U_{4,0}$ & $U_{4,1}$ \\
        \cline{2-3} 
           & $(\frac{3\pi}{2}, 3.8370, 1.9164)$ & $(7.853, 0.6599, 0.3456)$ \\
        \cline{2-3}
        \hline
    \end{tabular}
    \caption{Quantum strategies for four players game given by Eq. (\ref{eq:GHZ}) where the maximum advantage is achieved with a $\ket{GHZ}$ state. With the quantum
 strategy, the players win with a probability $\approx 0.8535$ compared to $0.6225$ with classical strategies. This demonstrates a significant advantage for quantum strategies, with a gap of 22.5\% between the quantum and classical success probabilities.}
    \label{tab:examples_GHZ_strategies}
\end{table}
\FloatBarrier

\subsection{Gain with $\ket{W}$}
In order to find the equation where $\ket{W}$ has a better performance than $\ket{GHZ}$, we need to "break" the symmetries of Eq. (\ref{eq:GHZ}). This is done again by generalizing to the four-qubit case an equation of \cite{Jaffali2024}. It is thus possible to find a four-qubit game where using a $\ket{W}$ state gives a better advantage compared to a $\ket{GHZ}$ state : 

\begin{equation}\label{eq:w-equation}
(w  x)+(w  y) + (w  z) + (x  y) + (x  z) + (y  z) = (\overline{a} b c d) + (a \overline{b} c d) + (a b \overline{c}d) + (a b c \overline{d}).
\end{equation}

The best classical strategy allows one to win this new game with probability $0.6875$ while the gain increases to probability approximately $0.7499$ when using a $\ket{W}$ state for the quantum strategy (see Table \ref{tab:examples_W_strategies}). In comparison, with a $\ket{GHZ}$ state, the success probability only reaches 
$0.5727$.
\begin{table}[h!] 
    \centering
    \begin{tabular}{|c|c|c|}
        \hline 
        Equation  & $U_{1,0}$ & $U_{1,1}$ \\
        \cline{ 1-2} 
        \hline
         & $(8.6486, 1.5440, 6.4701)$ & $(3.3477, 0.8709, 0.1882)$\\
        \cline{2-3}
        \cline{2-3}
        &  $U_{2,0}$ & $U_{2,1}$ \\
        \cline{2-3} 
        $(w  x)+(w  y) + (w  z) + (x  y) + (x  z) + (y  z)$ &  $(3.9180, 0.2608,3.3286)$ & $(3.3475, 2.7149, 0.1868)$ \\
        \cline{2-3} 
        \cline{2-3} 
       $ = (\overline{a} b c d) + (a \overline{b} c d) + (a b \overline{c}d) + (a b c \overline{d})$ & $U_{3,0}$ & $U_{3,1}$ \\
        \cline{2-3} 
           & $(3.9178, 5.6411, -2.9545)$ & $(2.9354, 5.5383, 3.3283)$ \\
        \cline{2-3}
        \cline{2-3} 
        \cline{2-3} 
        & $U_{4,0}$ & $U_{4,1}$ \\
        \cline{2-3} 
           & $(2.3654, 0.7975, 6.4703)$ & $(9.6308, 3.4886, 0.1862)$ \\
        \cline{2-3}
        \hline
    \end{tabular}
    \caption{Quantum strategies for four players game where the maximum advantage is achieved with a $\ket{W}$ state. With the quantum
 strategy, the players win with a probability $\approx 0.7499$ compared to $0.6875$ with classical strategies. This demonstrates a significant advantage for quantum strategies, with a gap of 6.25\% between the quantum and classical success probabilities.}
    \label{tab:examples_W_strategies}
\end{table}
\FloatBarrier

\subsection{Other four-qubit entangled states: the critical states}
In the four-qubit Hilbert space there are an infinite number of different classes of entanglement (see Section \ref{subsec:4qubit}) and the question of deciding what a maximally four-qubit entangled state is, is not fully answered \cite{enriquez2016maximally}. In this subsection one tests some alternative of four genuine entangled states known as critical states \cite{oeding2024}. Critical states maximize some entanglement monotones defined by algebraic invariants. Here are the explicit forms of three four-qubit critical states:
\begin{itemize}
\item $\ket{MP}=\dfrac{1}{2}(\ket{0000}+\ket{0011}+\ket{1100}+\ket{1111})$, known as the Mermin-
 Peres state.
 \item $\ket{C_1}=\dfrac{1}{2}(\ket{0000}+\ket{0011}+\ket{1100}-\ket{1111})$, the cluster state.
 \item $\ket{L}=\dfrac{1}{4}[(1+\omega)(\ket{0000}+\ket{1111})+(1-\omega)(\ket{0011}+\ket{1100})+\omega^2(\ket{0110}+\ket{1001}+\ket{1010}+\ket{0101})]$.
\end{itemize}
The $\ket{MP}$ state is used as a resource to play the quantum game known as the Memin-Peres Magic square but also variation of those games \cites{kelleher2023implementing}. Cluster states are states that maximize the Renyi $\alpha$-entropy\cite{gour2010all,enriquez2016maximally}, by symmetry of the role of the players one only considers $\ket{C_1}$, the other cluster states being obtained by transposition. Finally $\ket{L}$ is a state that maximizes the hyperdeterminant, an analogue of the $3$-tangle for four-qubit states, see \cites{gour2010all, chen2013proof}.

The next two tables summarize our findings for those three critical states. Table \ref{table:criticalEq9} provides the score of the best quantum strategy to play the game given by Eq. (\ref{eq:GHZ}). Then, Table \ref{table:bestcritical} gives the best possible quantum gain and the corresponding best classical  when we consider all games of type $f(x,y,z,x)=a\oplus b\oplus c\oplus d$.

\begin{table}[!h] 
    \centering
    \begin{tabular}{|c|c|c|}
        \hline 
        Equation & State & Best Quantum Gain \\
        \hline
        \cline{2-3}
        & $\ket{MP}$ &0.6767 \\
        \cline{2-3}
 $(x y  z) + (x  y \overline{w}) + (x  z  \overline{w}) + (y  z  \overline{w} )$        & $\ket{C_1}$ & 0.6767 \\
 \cline{2-3}
   $+(w  \overline{x} \overline{y} \overline{z}) = a \oplus b \oplus c \oplus d $      & $\ket{L}$ & 0.6767\\
   \hline
\end{tabular}
\caption{Best quantum gain when playing the four-qubit game of Eq. (\ref{eq:GHZ}) with the four-qubit critical states, $\ket{MP}$, $\ket{C_1}$ and $\ket{L}$. One recalls that the best classical gain for Eq. (\ref{eq:GHZ}) is $0.6225$.}\label{table:criticalEq9}
\end{table}

\begin{table}[!h] 
    \centering
    \begin{tabular}{|c|c|c|}
        \hline 
         State & Best Quantum Gain &Best classical \\
        \hline
         $\ket{MP}$ & 0.7499  & 0.625\\
        \hline
      $\ket{C_1}$ &0.7499 &0.625 \\
 \hline
 $\ket{L}$ & 0.6767 & 0.625\\
   \hline
\end{tabular}
\caption{Best quantum gains when we analyze all possible strategy for all possible games of type $f(x,y,z,w)=a\oplus b\oplus c\oplus d$.}\label{table:bestcritical}
\end{table}

\subsection{The four-qubit classification}\label{subsec:4qubit}
It is well-known, as demonstrated by \cite{verstraete2002four}, that there exist nine types of entangled four-qubit states. More precisely, the number of classes of entanglement under the so-called SLOCC group is infinite but can be grouped in $9$ families, $6$ of them depending on parameters, $3$ being parameters free. Let us recall with Table \ref{table:4qubit} the normal forms of those families (in its corrected version from \cite{chterental2006normal}).

In this section, various four-qubit entangled states are tested, with each representing one of these nine types. 
\begin{table}[!h] 
\centering
\begin{tabular}{|c|c|}
\hline
\textbf{Name} & \textbf{Normal Form} \\
\hline
${G_{abcd}}$ & $\frac{a+d}{2}(\ket{0000}+\ket{1111})+\frac{a-d}{2}(\ket{0011}+\ket{1100}\rangle)$\\
  & $+\frac{b+c}{2}(\ket{0101}+\ket{1010})+\frac{b-c}{2}(\ket{0110}+\ket{1001}\rangle)$ \\
\hline
${L_{abc_2}}$ & $\frac{a+b}{2}(\ket{0000}+\ket{1111})+\frac{a-b}{2}(\ket{0011}+\ket{1100})+c(\ket{1010}+\ket{0101})+\ket{0110}$ \\
\hline
${L_{a_2b_2}}$ & $a(|0000\rangle+|1111\rangle)+b(|0101\rangle+|1010\rangle)+|0110\rangle+|0011\rangle$\\
\hline
${L_{ab_3}}$ & $a(\ket{0000}+\ket{1111})+\frac{a+b}{2}(  \ket{0101}+\ket{1010})+\frac{a-b}{2}(  \ket{0110}+\ket{1001})$\\
 & $+\frac{i}{\sqrt{2}}(\ket{0001}+\ket{0010}-\ket{1110}-\ket{1101})$\\
\hline
${L_{a_4}}$ & $a(\ket{0000}+\ket{0101}+\ket{1010}+\ket{1111})+i\ket{0001}+\ket{0110}-i\ket{1011}$ \\
\hline
${L_{a_20_{3\oplus {1}}}}$ & $a(\ket{0000}+\ket{1111})+\ket{0011}+\ket{0101}+\ket{0110}$\\
\hline
\hline
${L_{0_{7\oplus \overline{1}}}}$  & $\ket{0000} + \ket{1011} + \ket{1101} + \ket{1110}$ \\
\hline
$L_{0_{5\oplus \overline{3}}}$  & $\ket{0000} + \ket{0101}+\ket{1000} + \ket{1110}$ \\
\hline
$L_{0_{3\oplus \overline{1}}0_{3\oplus \overline{1}}}$  & $\ket{0000} + \ket{0111}$ \\
\hline
\end{tabular}
\caption{The four-qubit classification into $9$ families of entanglement following \cite{verstraete2002four} and \cite{chterental2006normal}. Up to local changes of basis, i.e. SLOCC$=SL_2(\CC)\times SL_2(\CC)\times SL_2(\CC)\times SL_2(\CC)$ transformations, all four-qubit quantum states can be transformed into one of those 9 families. Note that the first $6$ depend on parameters.}\label{table:4qubit}
\end{table}

In Table \ref{tab:recap_fusion} we report the score obtained for the best quantum strategy when playing the game defined by Eq. (\ref{eq:GHZ}) with an instance of each of the nine-family. For the 6 families depending on parameters we randomly choose a set of parameters (we repeated it four times) and collected the best gain and also the average quantum gain over the four sets of parameters. Once again with Eq. (\ref{eq:GHZ}) we do not get any higher score than $0.8535$.

\begin{table}[h!] 
    \centering
    \begin{tabular}{|c|c|c|c|}
        \hline 
        Equation & State & Best Quantum Gain & Average Gain \\
        \hline
        \cline{2-4}
        & $G_{abcd}$ & 0.8535 & 0.7737 \\
        \cline{2-4}
        & $L_{abc_2}$ & 0.8532 & 0.7524 \\
        \cline{2-4}
        & $L_{a_2b_2}$ & 0.8534 & 0.7250 \\
        \cline{2-4}
  $(x y  z) + (x  y \overline{w}) + (x  z  \overline{w}) + (y  z  \overline{w} )$       & $L_{ab_3}$ & 0.7357 & 0.7072 \\
        \cline{2-4}
    $+(w  \overline{x} \overline{y} \overline{z}) = a \oplus b \oplus c \oplus d $     & $L_{a_4}$ & 0.6985 & 0.6774 \\
         \cline{2-4}
        & $L_{a_20_{3\bigoplus 1}}$ & 0.8535 & 0.7164 \\
        
        \cline{2-4} 
        & $L_{0_{7\bigoplus \overline{1}}}$ & 0.6586 & X \\
        \cline{2-4} 
        & $L_{0_{5\bigoplus \overline{3}}}$ & 0.6530 & X \\
         \cline{2-4} 
        & $L_{0_{3\bigoplus \overline{1}}0_{3\bigoplus \overline{1}}}$ & 0.7499 & X \\
        \hline
    \end{tabular}
    \caption{Quantum gain for different parametric and non-parametric four-qubit entangled states when playing the four-player games defined by Eq. (\ref{eq:GHZ}). The first six families of the four-qubit classification depend on parameters and we have computed the best quantum strategies for different random choices of parameters (we randomly repeated it four times). The first column provides the best quantum advantage obtained and the second one the average best gain over the different random choices of parameters. This does not apply to the last three classes that are parameter-free. Recall that the classical gain for Eq. (\ref{eq:GHZ}) is $0.6225$. With a $\ket{GHZ}$ states one can win the game with probability $0.8535$ and if one uses $\ket{W}$ as our entangled resource the gain is of $0.69$.}
    \label{tab:recap_fusion}
\end{table}
\FloatBarrier

If one now considers all games based on $f(x,y,z,w)=a\oplus b\oplus c\oplus d$ we do not get anything better than the difference between the best quantum strategy and the best classical ones that are reached for $\ket{GHZ}$.

To conclude this section, one would like to analyze how often a given family of four-qubit states provides a quantum advantage.

\subsection{Notion of game score}
 We define the notion of \textit{game score} for a given four-qubit entangled state and a given function $g(a,b,c,d)$ as the proportion in percentage of functions $f(w,x,y,z)$ which have a quantum gain higher than 1\% compare to the classical gain. \\

\begin{table}[h!] 
    \centering
    \begin{tabular}{|c|c|}
        \hline 
        State & Game Score \\
        \cline{1-2} 
        \hline
        $G_{abcd}$ & 0.3452 \\
        \cline{1-2}
        $L_{abc_2}$ & 0.2622 \\
        \cline{1-2}
        $L_{a_2b_2}$ & 0.2544 \\
        \cline{1-2}
        $L_{ab_3}$ & 0.2450 \\
        \cline{1-2}
        $L_{a_4}$ & 0.2179 \\
        \cline{1-2}
        $L_{a_20_{3\bigoplus 1}}$ & 0.2554 \\
        \cline{1-2}
        $L_{0_{7\bigoplus \overline{1}}}$ & 0.0031 \\
        \cline{1-2}
        $L_{0_{5\bigoplus \overline{3}}}$ & 0.0033 \\
        \cline{1-2}
        $L_{0_{3\bigoplus \overline{1}}0_{3\bigoplus \overline{1}}}$ & 0.2804 \\
        \hline
    \end{tabular}
    \caption{Game scores for different four-qubit entangled states and $g(a,b,c,d)=a\oplus b\oplus c\oplus d$.}
    \label{tab:recap_fusion_ordered_with_W}
\end{table}
\FloatBarrier

In the case where $\ket{GHZ}$ is used and $g(a,b,c,d)=a \oplus b \oplus c \oplus d $, a game score of $26.34\%$ is obtained. In other words, on the $3907$ functions $f$ to test, $26.34\%$ of them have a quantum gain higher than 1\% compared to the classic one. Concerning the average gap between quantum and classic gain, in this case, it is equal to $5.49\%$. When we consider the $\ket{W}$ state this game score drops to $24\%$

Table \ref{tab:recap_fusion_ordered_with_W} collects the game score obtained for the nine types of entangled four-qubit states. For parametric states, one randomly chose the parameters and took the best game score after several trials. When a random $G_{abcd}$ state is used, a game score of 34.52\% is obtained, much better than with a $\ket{GHZ}$ state. But concerning the average gap, this one is equal to  3.73\%, so lower than the previous one with $\ket{GHZ}$. \\


\subsection{Retrieving  gains of Table \ref{tab:recap_fusion}}
We now explain why, in Table \ref{tab:recap_fusion}, one recovers for the families $G_{abcd}, L_{abc_2}$ $L_{a_2b_2}$ and $L_{a_{2}0_{3\bigoplus 1}}$ the same best quantum gain as the one obtained with a $\ket{GHZ}$ state. We also discuss the quantum gain for  $L_{ab_3}$ and $L_{a_4}$ and their relation to the critical states $\ket{MP}$.

\subsubsection{Retrieving $\ket{GHZ}$ Gain}
Regarding the parametric four-qubit entangled states, achieving the same gain as $\ket{GHZ}$ is possible when those parametric states can yield a $\ket{GHZ}$ state. It appears clearly that $\ket{GHZ}$ belongs to an orbit of the family $G_{abcd}$. For instance, one can choose $a=b$ and $c=d=0$ to recover the usual $\ket{GHZ}$ state. 

If one considers the family $L_{abc_2}$ with $c=0$ one has 
\[ \lim_{\substack{a=b \to \infty}} L_{abc_2} = \ket{GHZ}.\] 

And the same holds if $a=\pm b\to \pm\infty$. Therefore:


\begin{equation}\label{eq:limits} \lim_{\substack{a=\pm b \to \pm\infty \\ c=0}} Gain( L_{abc_2}) = Gain(\ket{GHZ}) = 0.8535. \end{equation}

\begin{figure}[h]
    \centering
    \includegraphics[width=0.4\linewidth]{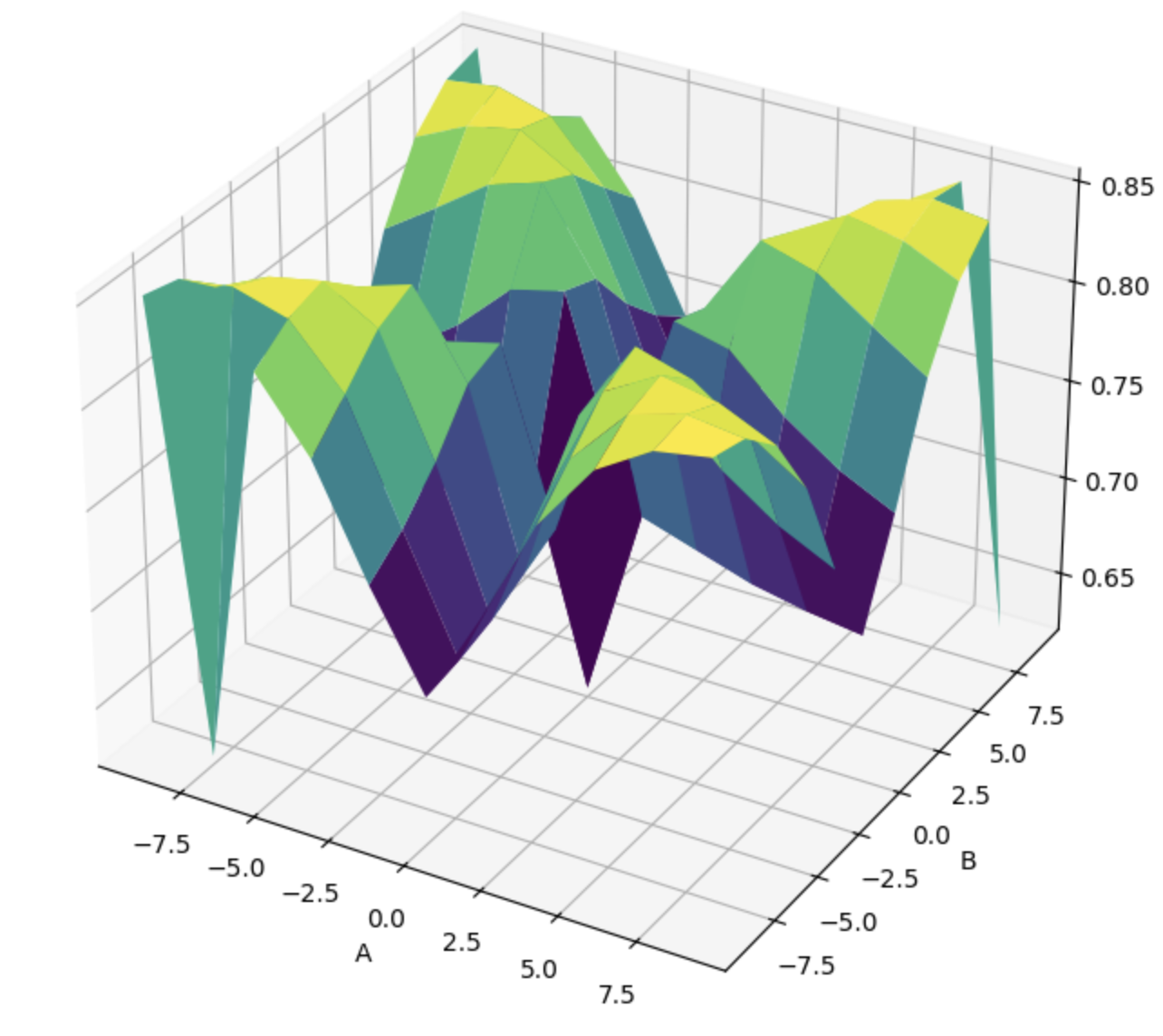}
    \includegraphics[width=0.4\linewidth]{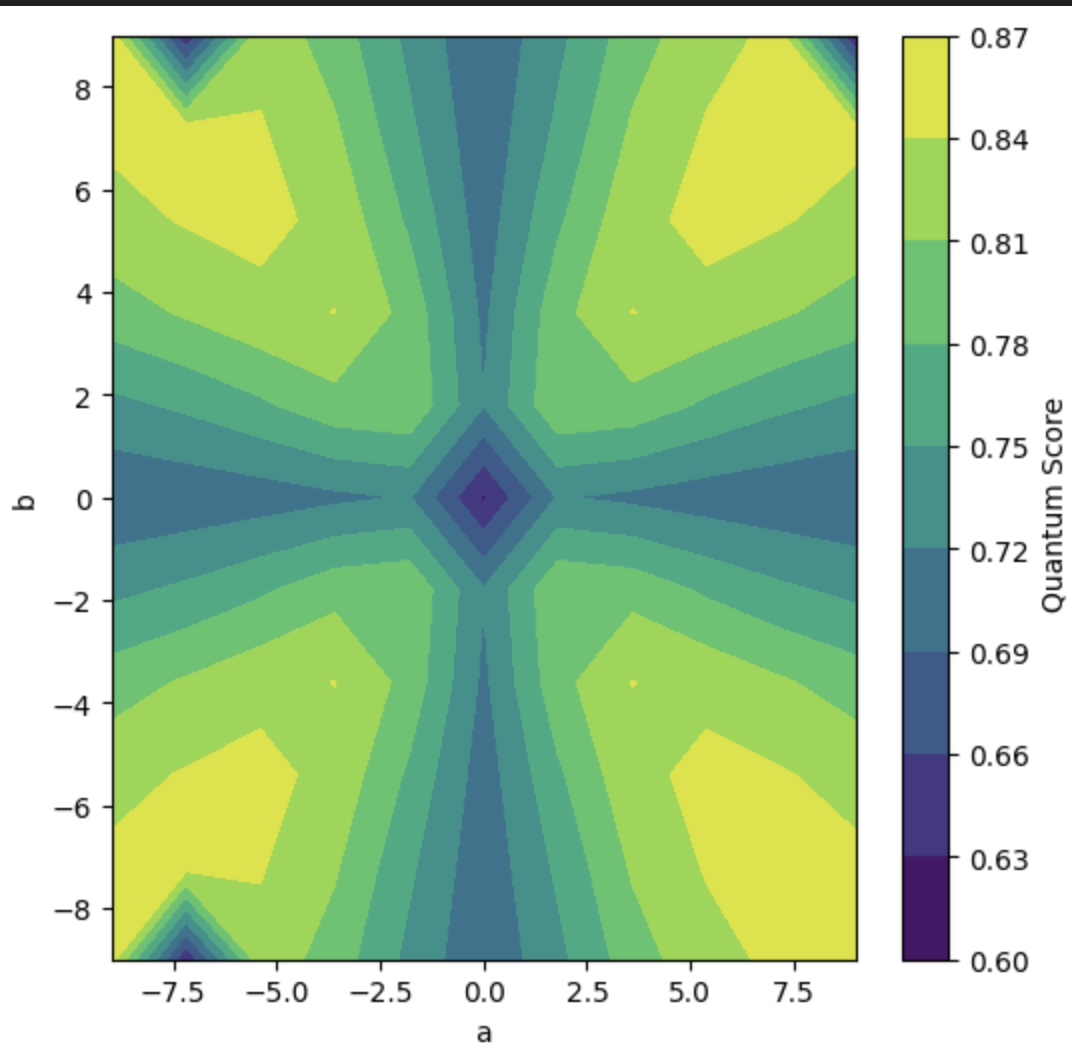}
    \caption{Evolution of the quantum gain of $L_{abc_2}$ with $c=0$ for Eq. (\ref{eq:GHZ}) with $a$ and $b$ between $-9$ and $9$. 2D contour plot of the Quantum Score of $L_{abc_2}$ with $c=0$ for Eq. (\ref{eq:GHZ}) with $a$ and $b$ between $-9$ and $9$.}
    \label{fig:labc2}
\end{figure}

To illustrate this property one plotted in Figure \ref{fig:labc2} the gain for the game based on Eq. (\ref{eq:GHZ}) with the quantum state $L_{abc_2}$ for different values of $a$ and $b$ and $c_2=0$. The limits of Eq. (\ref{eq:limits}) can be recovered from Figure \ref{fig:labc2}.

With \( L_{a_2b_2} =  a (\ket{0000} + \ket{1111}) + b (\ket{0101} + \ket{1010}) + \ket{0110} + \ket{0011} \), two \(\ket{GHZ}\) states can be observed: the first is defined by \( a (\ket{0000} + \ket{1111}) \) and the second is defined by \( b (\ket{0101} + \ket{1010}) \). So by studying the quantum gain of $ L_{a_2b_2}$ when $a$ goes to $\infty$ and $b$ goes to $0$, it is possible to deduce that :

\[ \lim_{\substack{a \to \infty \\ b \to 0}} Gain( L_{a_2b_2}) = \lim_{\substack{b \to \infty \\ a \to 0}} Gain( L_{a_2b_2})= Gain(\ket{GHZ}) = 0.8535 \] 

\begin{figure}[h]
    \centering
    \includegraphics[width=0.4\linewidth]{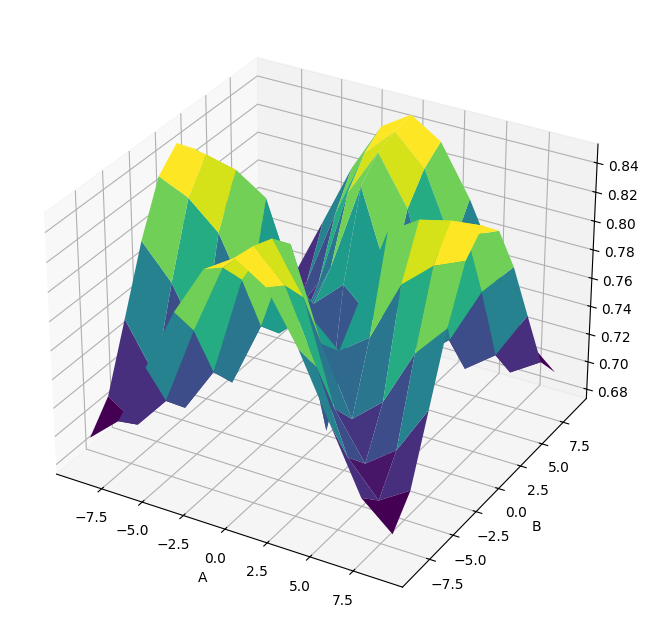}
    \includegraphics[width=0.4\linewidth]{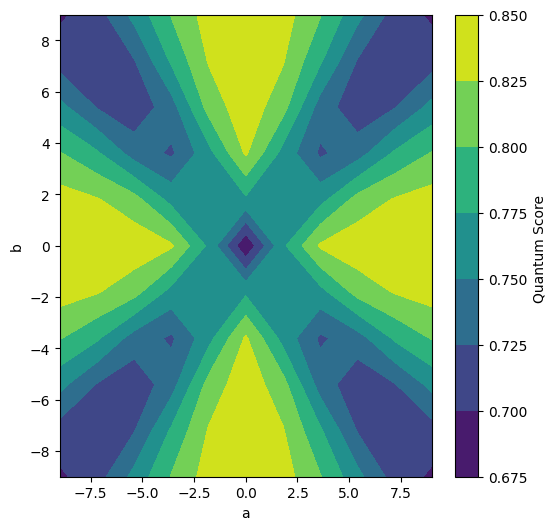}
    \caption{Evolution of the quantum gain of $L_{a_2 b_2}$ for Eq. (\ref{eq:GHZ}) with $a$ and $b$ between $-9$ and $9$. 2D contour plot of the Quantum Score of $L_{a_2b_2}$ on the Eq. (\ref{eq:GHZ}) with $a$ and $b$ between -9 and 9.}
    \label{fig:lab2}
\end{figure}

We can clearly see this in Figure \ref{fig:lab2}, we can also observe that the quantum gain is minimum when $a=b$.

In the case of $L_{a_{2}0_{3\bigoplus 1}} = a (\ket{0000}+\ket{1111}) +(\ket{0011}+\ket{0101}+\ket{0110})$
the $\ket{GHZ}$ state is again clearly visible. Thus, it is possible to retrieve $\ket{GHZ}$ score, for $(x y  z) + (x  y \overline{w}) + (x  z  \overline{w}) + (y  z  \overline{w} ) + (w  \overline{x} \overline{y} \overline{z}) = a \oplus b \oplus c \oplus d $ by increasing the coefficient $a$.  One observes that : 
\[ \lim_{a\to\infty} Gain(L_{a_{2}0_{3\bigoplus 1}}) = Gain(\ket{GHZ}) = 0.8535 \]. 

\begin{figure}[ht]
    \centering
    \includegraphics[width=0.55\linewidth]{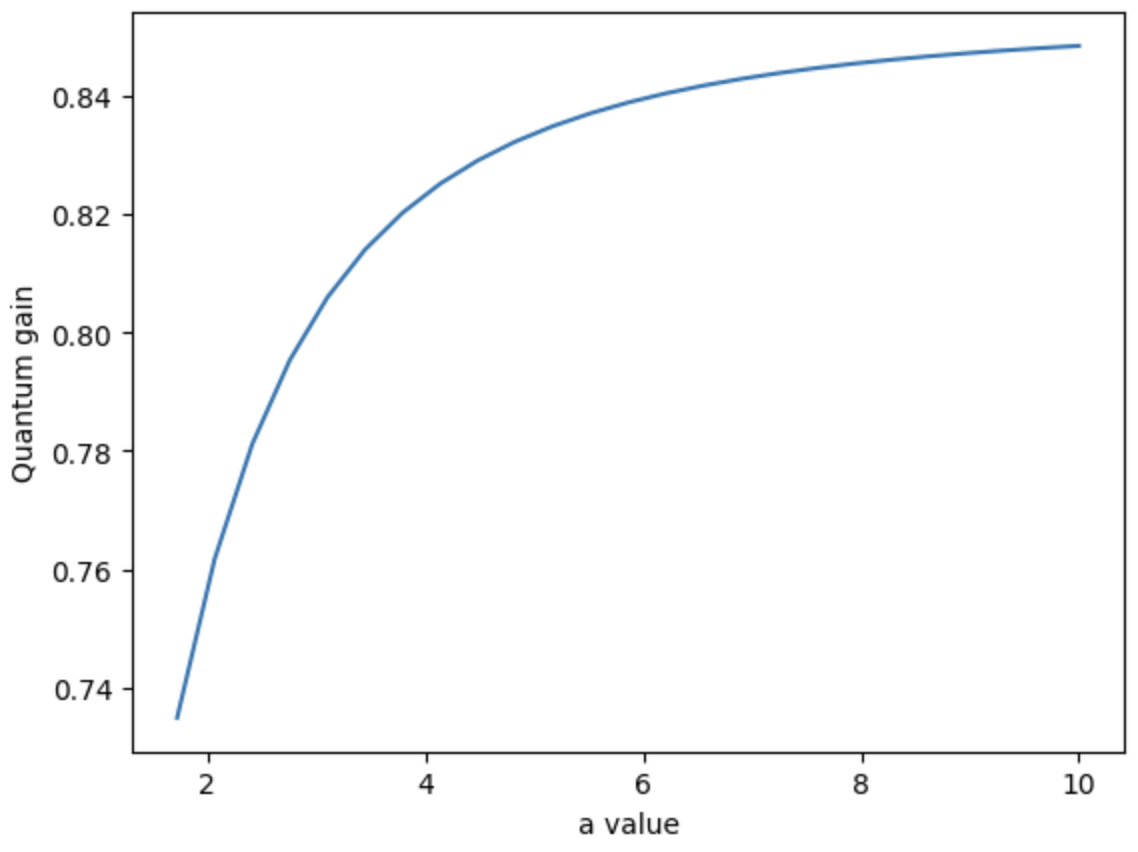}
    \caption{Evolution of the Quantum of the Gain of $L_{a_20_{3\bigoplus 1}}$ with $a$ increasing from 2 to 10 approaches $\ket{GHZ}$ state.}
    \label{fig:la203}
\end{figure}
\FloatBarrier

\subsubsection{Retrieving $\ket{MP}$ Gain}
Let us now consider the parametric states $L_{ab_3}$ and $L_{a_4}$ when used with Eq. (\ref{eq:GHZ}).
 We first observe that $\ket{MP}$ belongs to the  $L_{ab_3}$ family for $a=b=0$. Indeed we have: 
\[L_{ab_3}\underbrace{=}_{a=b=0}\frac{i}{\sqrt{2}}(\ket{0001}+\ket{0010}-\ket{1110}-\ket{1101}=\frac{i}{\sqrt{2}}((\ket{00}-\ket{11})(\ket{10}+\ket{01}).\]

\begin{figure}[!h]
    \centering
    \includegraphics[width=0.4\linewidth]{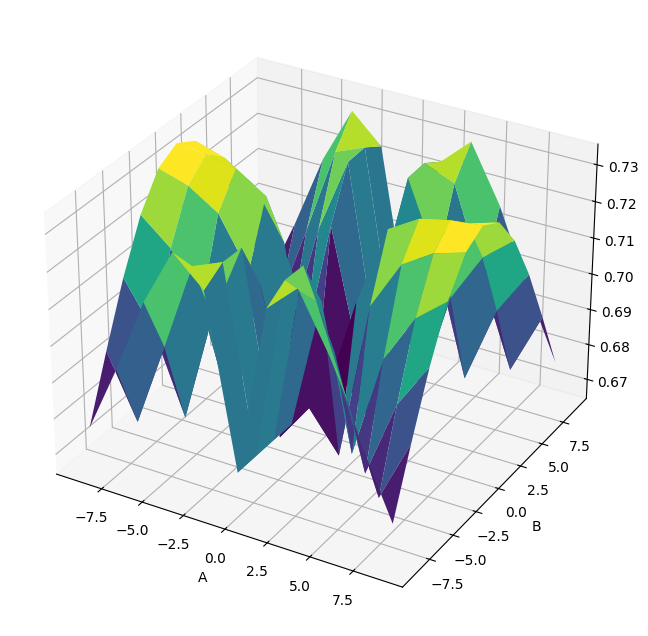}
    \includegraphics[width=0.4\linewidth]{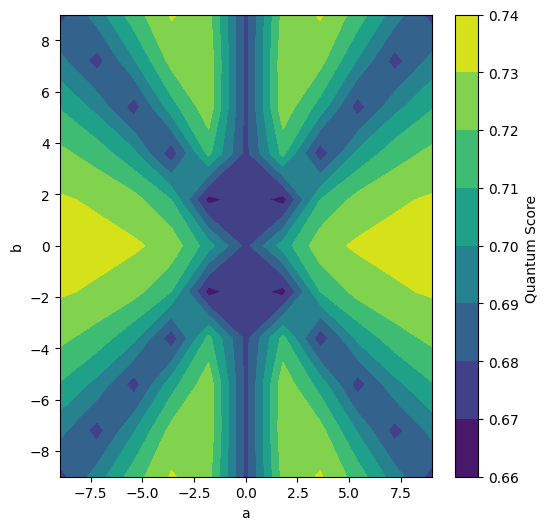}
    \caption{Evolution of the quantum gain of $L_{ab_3}$ for Eq. (\ref{eq:GHZ}) with $a$ and $b$ between $-9$ and $9$. 2D contour plot of the Quantum Score of $L_{ab_3}$ for Eq. (\ref{eq:GHZ}) with $a$ and $b$ between -9 and 9.}
    \label{fig:Lab3}
\end{figure}

Therefore as one can see with Figure \ref{fig:Lab3}:
\[ \lim_{\substack{a \to 0 \\ b \to 0}} Gain( L_{ab_3}) = Gain(\ket{MP}) = 0.6767\]

\begin{figure}[!h]
    \centering
    \includegraphics[width=0.7\linewidth]{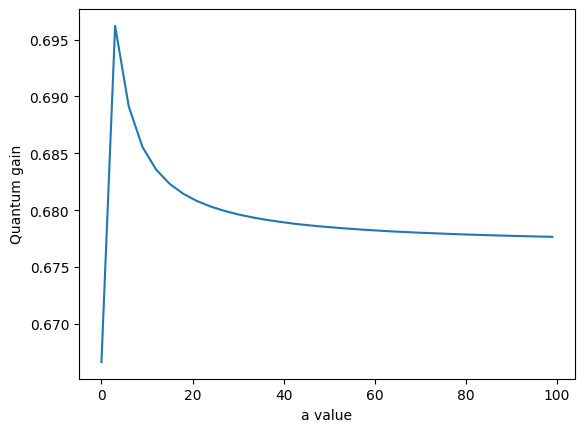}
    \caption{Evolution of the Quantum  Gain of $L_{a_4}$ for Eq. (\ref{eq:GHZ}) with $a_4$ increasing from $0$ to $100$. The gain approaches the gain obtained with the critical state $\ket{MP}$.}\label{La4}
\end{figure}

Similarly, up to a permutation of the second and third qubit, one can see that 
$$\lim_{a\to \infty} L_{a_4}=\ket{MP}$$
Figure \ref{La4} illustrates this fact as one sees the gain to converge to $0.6767$ when $a$ increases.
\[ \lim_{a\to \infty} Gain( L_{a_4}) = Gain(\ket{MP}) = 0.67677.\]

\section{Four-qubit CHSH games on a quantum computer} \label{sec:ibm}
In this section, two different four-qubit CHSH games are played on real quantum devices. The first objective is to observe the violation of classical bounds through our adapted CHSH games, while the second is to evaluate the performance of the quantum protocol in a noisy environment and on an actual quantum device. Two distinct games are selected: one where the $|GHZ\rangle$ state outperforms the classical gain, and another where the $|W\rangle$ state surpasses both the classical limit and the $|GHZ\rangle$ state in terms of score. Those are the games studied in Section \ref{sec:4qgame}. In both scenarios, the experimental advantage given by the quantum strategies surpass the classical bound as predicted by the theory.

\subsection{Experiment with $\ket{GHZ}$}

We first test the game defined by Eq. (\ref{eq:GHZ}). Recall that, in that case, one can outperform the best classical strategy ($0.6225$ of gain) with a quantum strategy given in Table \ref{tab:examples_GHZ_strategies} involving the $\ket{GHZ}$ state ($0.8535$ of gain).

\begin{figure}[!h]
    \centerline{
    \Qcircuit @C=1em @R=.7em {
    & \gate{H} & \ctrl{1} & \qw & \qw & \qw & \gate{U_{1,w}} & \meter & \qw\\
    & \qw & \targ  & \ctrl{1} & \qw & \qw & \gate{U_{2,x}} & \meter & \qw\\
    & \qw & \qw  & \targ & \ctrl{1} & \qw  & \gate{U_{3,y}} & \meter & \qw\\
    & \qw & \qw  & \qw & \targ & \qw  & \gate{U_{4,z}} & \meter & \qw\\
    }}
    \caption{Quantum circuit used for applying the quantum strategy with $\ket{GHZ}$. The optimal angles of the local unitary transformations depend on the question sent to the players and are
reported in Table \ref{tab:examples_GHZ_strategies}.}
    \label{fig:ghz_state}
\end{figure}
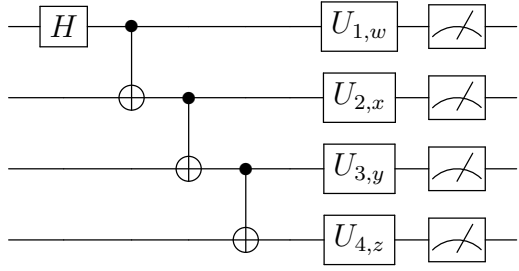

To play the game, the $\ket{GHZ}$ state is first generated by applying a Hadamard gate on the first qubit followed by a series of CNOT gates. Subsequently, each player applies a unitary transformation on their qubit based on the received question, with angles provided in Table \ref{tab:examples_GHZ_strategies}. Finally, each qubit is measured in the computational basis, and the results are returned to the referee. The circuit executed for this process is depicted in Figure \ref{fig:ghz_state}.

\begin{figure}[!h]
    \centering
    \includegraphics[width=0.8\linewidth]{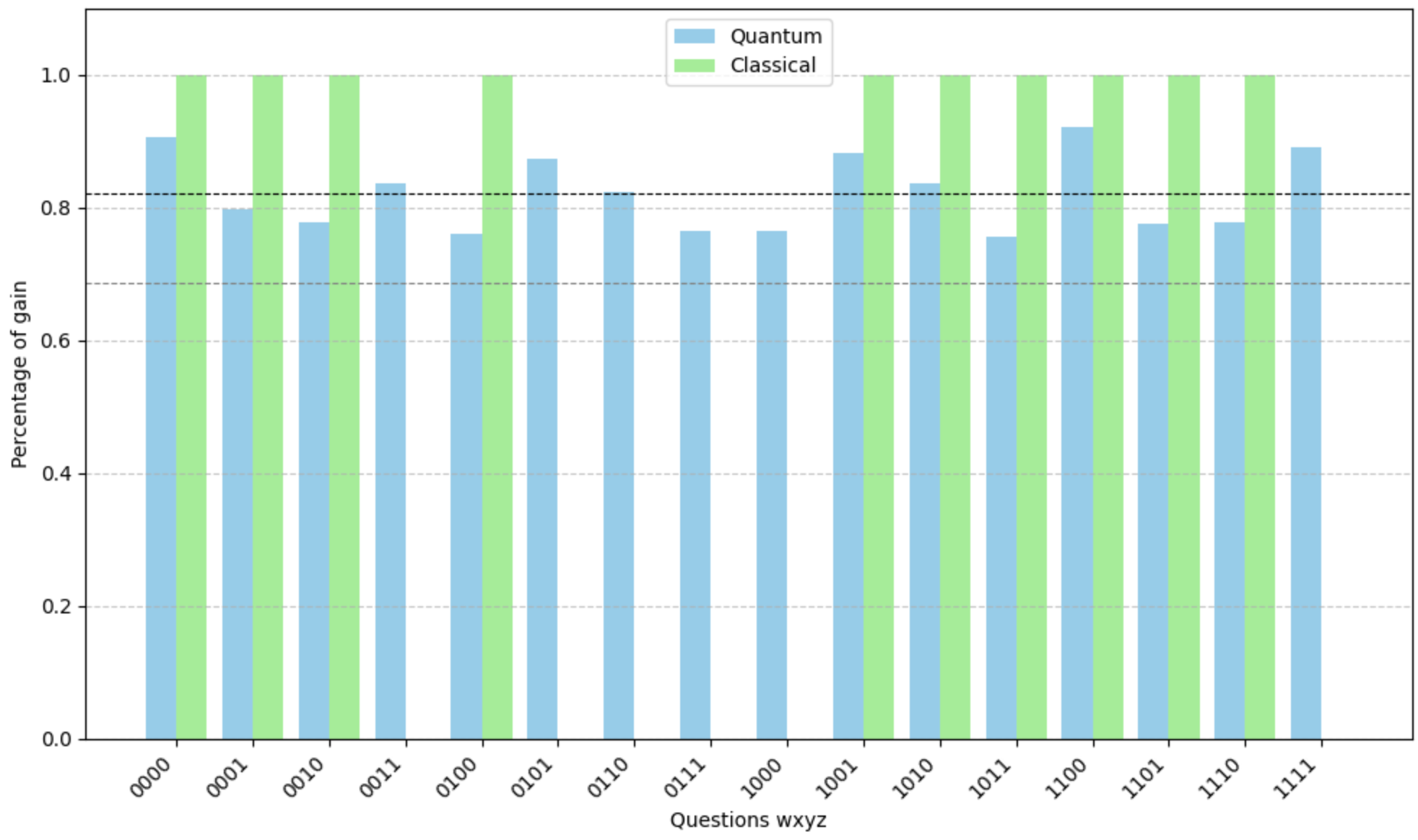}
    \caption{Histogram of the percentage of victory as a function of the question sent to the four players in the game defined by Eq. (\ref{eq:GHZ}). In green a classical strategy that achieves an average gain of 62.25\% of victory. In blue, the quantum gain with an average score of 82.19\%, the four players sharing a $\ket{GHZ}$ entangled state. The dashed line represents the average victory. The
experiment was performed on IBM Sherbrooke with 10,000 shots on October 14, 2024.}
    \label{fig:game-ghz}
\end{figure}
\FloatBarrier

\subsection{Experiment with $\ket{W}$}

For the second experiment, one considers the game where the four players win whenever their answers $(a,b,c,d)\in \mathbb{B}^4$ satisfy Eq. (\ref{eq:w-equation}) for a choice of question $(w,x,y,z)\in \mathbb{B}^4$. \\
In order to construct a $\ket{W}$ state with four-qubits we use a method inspired by \cite{Diker:2022}, and adapted by \cite{decoodt_2018} using rotations around the $Y$-axis, controlled-$Z$, controlled-$NOT$ and Hadamard gates:



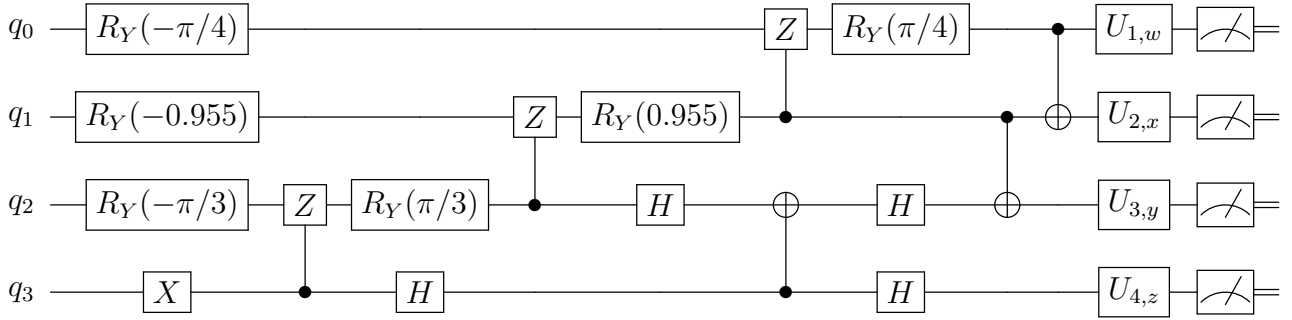
\begin{figure}[!h]
    \centerline{
    \Qcircuit @C=0.8em @R=1.2em {
    \lstick{q_0} & \gate{R_Y(-\pi/4)} & \qw & \qw       & \qw      & \qw      & \gate{Z}     & \gate{R_Y(\pi/4)} & \qw  &\ctrl{1} &\gate{U_{1,w}}& \meter & \cw \\
    \lstick{q_1} & \gate{R_Y(-0.955)} & \qw   &\qw      &  \gate{Z}    & \gate{R_Y(0.955)} & \ctrl{-1} & \qw&\ctrl{1} &\targ    & \gate{U_{2,x}}&\meter & \cw \\
    \lstick{q_2} & \gate{R_Y(-\pi/3)} & \gate{Z} & \gate{R_Y(\pi/3)} &\ctrl{-1}& \gate{H}         & \targ &\gate{H}    & \targ &\qw     & \gate{U_{3,y}}&\meter & \cw \\
    \lstick{q_3} & \gate{X}           & \ctrl{-1}    & \gate{H} & \qw      & \qw      & \ctrl{-1}      & \gate{H} & \qw &\qw& \gate{U_{4,z}}&\meter & \cw \\}}
    \caption{Quantum circuit used for applying the quantum strategy with $\ket{W}$. The optimal angles of the local unitary transformations depend on the question sent to the players and are
reported in Table \ref{tab:examples_W_strategies}.}
    \label{fig:w_state}
\end{figure}
\FloatBarrier

\begin{figure}[!h]
    \centering
    \includegraphics[width=0.85\linewidth]{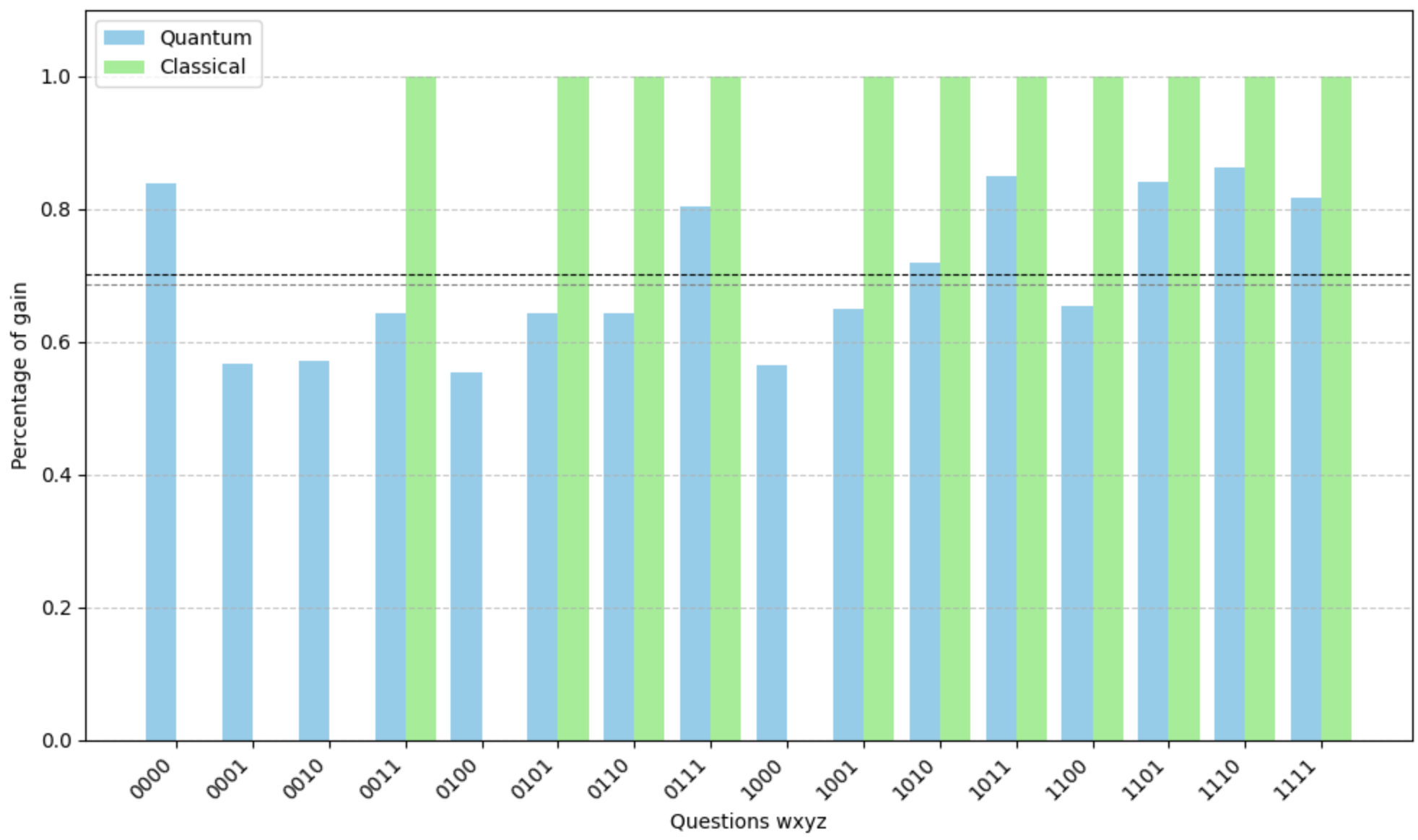}
    \caption{Histogram of the percentage of victory as a function of the question sent to the four players for the game defined by Eq. (\ref{eq:w-equation}). In green a classical strategy that achieves an average gain of 68.75\% of victory. In blue, the quantum gain with an average score of 70.13\%, the four players sharing a $\ket{W}$ entangled state. The dashed line represents the average victory. The
experiment was performed on IBM Sherbrooke with 10,000 shots on October 14, 2024.}
    \label{fig:w-result}
\end{figure}

We ran this game on IBM\_Sherbrooke device. Figure \ref{fig:w-result} presents our results in the form of a histogram comparing the percentage of victory when playing the game with a $\ket{W}$ state and with the best classical strategy. On average, the percentage of wins is higher with $\ket{W}$ and beats the classical bound of $68.75\%$.

\FloatBarrier

\section{Conclusion} \label{sec:conclusion}

In this article, we explored the generalization of CHSH quantum games to four qubits, inspired by previous advances in two- and three-qubit games \cite{clauser1969proposed,Jaffali2024}. Our investigation led to the discovery and analysis of new quantum games. We explicitly described one game utilizing the four-qubit entangled state $\ket{GHZ}$ as the optimal resource and one game using the state $\ket{W}$ as the optimal resource. Significantly, we demonstrated that the usual quantum advantage of 10\%, observed in two- and three-qubit games, was not only achieved but also doubled, reaching 22.5\% in the case of the $\ket{GHZ}$ game. Additionally, we tested critical states that are candidates for maximally entangled states as well as generic states from the four-qubit classification \cite{verstraete2002four}, further expanding our understanding of the potential use of those different types of entanglement.

This advancement highlights not only the power of four-qubit entangled states to overcome classical limits but also the diversity of new possible games in this framework.  It is important to note that, despite the numerous results obtained, we had to reduce the problem size due to computational time constraints, making an exhaustive search impractical. However, our work confirms that for games based on conditions of types \[f(w,x,y,,z)=a\oplus b\oplus c\oplus d,\] 
the best quantum advantage, over all possible $f$ is obtained for the $\ket{GHZ}$ state. One expects this to remain true for a larger number of qubits. For practical applications, our framework offers a general method to decide if a four-player non-communication game, defined by a wining condition of type $f(w,x,y,z)=g(a,b,c,d)$, for $f$ and $g$ boolean functions, can be won advantageously with quantum entanglement. We hope this to provide in a near future scenario of quantum protocols for for instance self testing or device-independent quantum computing.

\section*{Codes and resources}
All codes and results of our quantum experiments are available at:
\url{https://github.com/ColibrITD-SAS/publications_material/tree/main/four_qubit_CHSH_game}
\bibliography{references}

\end{document}